\documentclass[utf8]{frontiersSCNS}
\usepackage{url,hyperref,lineno,microtype,subcaption}
\usepackage[onehalfspacing]{setspace}
\usepackage{listings}

\def\keyFont{\fontsize{8}{11}\helveticabold }
\def\firstAuthorLast{De Nicola {et~al.}} 
\def\Authors{Rocco De Nicola\,$^{1}$, Luca Di Stefano\,$^{2,*}$ and Omar Inverso\,$^{2}$}
\def\Address{$^{1}$IMT School for Advanced Studies, Lucca, Italy\\
$^{2}$Gran Sasso Science Institute (GSSI), L'Aquila, Italy  }
\def\corrAuthor{Luca Di Stefano}

\def\corrEmail{luca.distefano@gssi.it}

\let\footnotesize\scriptsize
\usepackage{xstring}
\usepackage{graphicx}
\usepackage{multicol}
\usepackage{alltt}
\usepackage{mathtools}
\usepackage{multicol}

\usepackage{tikz}
\usetikzlibrary{positioning,shapes}

\newcommand{\Put}[2]{\textsf{\textbf{put}}\left(#1\,\right)@#2}
\newcommand{\Get}[2]{\textsf{\textbf{get}}\left(#1\,\right)@#2}
\newcommand{\Qry}[2]{\textsf{\textbf{qry}}\left(#1\,\right)@#2}

\newcommand{\Self}{\mathsf{self}}
\newcommand{\This}{\mathsf{this}}

\newcommand{\Nil}{\mathbf{nil}}

\newcommand{\Quote}[1]{\textit{``#1''}}

\newcommand{\Circle}[1]{\IfEqCase{#1}{{white}{\mbox{\raisebox{0.3ex}{$\bigcirc$}}}}[\mbox{\color{#1}\Huge\raisebox{-0.3ex}{$\bullet$}}]}

\begin{document}
\onecolumn
\firstpage{1}

\title[Towards formal models and languages for verifiable Multi-Robot Systems]{Towards formal models and languages for verifiable Multi-Robot Systems} 

\author[\firstAuthorLast ]{\Authors} 
\address{} 
\correspondance{} 

\extraAuth{}

\maketitle

\begin{abstract}

\section{}
Incorrect operations of a Multi-Robot System (MRS) may not only lead to 
unsatisfactory results, but can also cause economic losses and threats to
safety. These threats may not always be apparent, since they may arise as
unforeseen consequences of the interactions between elements of the system.
This call for tools and techniques that can help in providing guarantees about MRSs behaviour.
We think that, whenever possible, these guarantees should be backed up by formal
proofs to complement traditional approaches based on testing and simulation.

We believe that tailored linguistic support to specify MRSs is a major step 
towards this goal.  In particular, reducing the gap between typical features of an MRS and 
the level of abstraction of the linguistic primitives would simplify both the
specification of these systems and the verification of their properties.
In this work, we review different agent-oriented languages and their features;
we then consider a selection of case studies of interest and implement them useing
the surveyed languages.  We also evaluate and compare effectiveness of the proposed solution, 
considering, in particular, easiness of expressing non-trivial behaviour.

\tiny
 \keyFont{ \section{Keywords:} multi-robot systems, languages, communication, 
 collective behaviour, automated reasoning}
\end{abstract}


\section{Introduction}
Multi-robot systems (MRSs) are an increasingly popular topic in robotics research.
Their broad range of activities have been categorised in several different ways~\citep{Brambilla2013,Arai2002,Bayndr2016}.
Typical tasks include exploration or patrolling, object transport and
manipulation (e.g. foraging), deployment (e.g. pattern formation), collective
decision making (e.g. flocking), task allocation, and many others.

Cooperation is the real power of a MRS:
by working together, the robots can globally achieve goals that would be
``difficult, if not impossible, to be accomplished by an individual robot''~\citep{Arai2002}.
On the other hand,
the concerns that typically arise with any robotic system~\citep{Vasic2013}
are largely exacerbated in the presence of multiple cooperating units.
In particular, the risk of incorrect operation, whence possible economic losses and even threats to safety, is much greater.
Such concerns are related to the inherent features of MRSs rather than the specific kind of task of a MRS.

A first source of trouble is 
\emph{open-endedness} of MRSs, i.e.
the fact that robots can dynamically enter or leave the system.
This happens when decommissioning faulty units, or deploying extra units
to increase the throughput or system's fault tolerance.
Another complication is \emph{anonymity}, in the sense that
cooperating robots may not necessarily rely on, or be aware of, each other's identity.
The identity is, for instance, irrelevant in a flock of drones that adjust their directions by looking at each other.
Anonymity has a particularly disruptive impact on communication,
as it renders existing traditional mechanisms such as point-to-point communication fundamentally inadequate.
Another cause of concern with MRSs is
\emph{decentralisation}, i.e. the absence of a central entity to coordinate the robots' activities.
Decentralisation makes synchronisation especially challenging, if possible at all,
and in general makes it difficult to achieve an acceptable robustness of the interaction protocols.
Another source of complexity in MRSs is the typically \emph{large size} of the 
systems, and in particular the considerably large state space
resulting from asynchronous interaction of large numbers of components. 

These features result in challenges to the specification of MRSs.
A vast portion of literature relies on general-purpose languages to model
MRSs~\citep{Pitonakova2013,Pitonakova2016,BuchananDynamicTaskPartitioning2016} 
but, typically, each work focuses on rather narrow classes of systems while 
making specific assumptions on the operating contexts.
In general, modelling the mentioned features of MRSs using general-purpose languages is not intuitive,
and can lead to increased code complexity,
higher likelihood of programming errors, and generally makes 
programs hard to develop and maintain, and complicates reasoning about them.

On the other hand, domain-specific languages with tailored, higher-level
primitives that reduce the conceptual gap with the above mentioned features can make
the specification of new MRSs easier, as well as constraining the complexity of
the resulting global behaviour~\citep{Mataric:1993:DEB:171174.171225}.
However, the heterogeneity of the domains might be detrimental for compositionality;
it may become harder to specify complex systems
by composing available solutions.
Also, there is some risk in adopting abstractions that are too specific,
as they might oversimplify the problem space and might not be able to realistically
describe scenarios of interest.
Thus, linguistic support should aim at addresing the aforementioned sources of
complexity, while achieving an acceptable trade-off between expressiveness and generality.

The nature of MRSs also makes their analysis problematic.
For instance, the correctness and efficiency of an MRS is commonly measured
through simulations or experiments in a real, yet controlled, environment.
However, interaction between components may give rise to complex collective 
behaviour that is difficult to
predict~\citep{MataricIssuesapproachesdesign1995}.
The interleaving of individual processes also means that it is often impossible
to systematically explore any possible behaviour through simulation:
subtle corner cases can go unnoticed and lead to failure in a real-world
deployment.
Languages equipped with a clear semantics can address these concerns by
supporting both informal reasoning and formal verification of properties.
Appropriate primitives can also prove helpful in this regard, since
verification can often exploit high-level information on the system to guide the analysis~\citep{Clarke1996,Flanagan2005}.
This, in turn, can make research on more complex systems feasible.
In the long term, such languages could become the core element of integrated
environments aiding the design of MRSs through automated reasoning tools,
quite like the currently available integrated development environments (IDEs) for 
programming languages.

In this paper we review some languages stemming from MRS and multi-agent systems
(MAS) literature.
Our selection is driven by their different nature and goals.
Buzz is oriented to real-world applications and shares some similarities with
popular languages, such as Python, JavaScript and Lua.
ISPL and its surrounding framework are specifically designed to enable reasoning
on the knowledge of agents, and provides explicit abstractions for the external
environment.
Finally, SCEL is a process description language where MRS features, such as as anonymity and
open-endedness, are transparent to the designer, allowing for a high degree of
naturalness in the specification of individual behaviour.
We highlight the main characterizing features of these three formalisms and 
compare them by considering how they model several traits commonly found in MRSs.
We also analyse how well the languages can support the design phase through
simulation or verification tools.
We focus on languages based on \emph{individual behaviour design}. 
That is, a developer using said languages is mainly concerned with the behaviour
of single components: the expected global behaviour is not explicitly programmed,
but is expected to arise from interaction between the robots.
Although alternative methods have been proposed, such as automatic design of 
individual behaviour from higher-level
specifications~\citep{Ulusoy2013,Nikou2016} or top-down behavioural
languages~\citep{Bachrach2008}, the bottom-up approach is still
widely adopted due to its intuitiveness~\citep{Brambilla2013}.


To guide our comparison, we focus on two popular case studies, namely \emph{foraging} and \emph{flocking}.
When foraging, robots must find items in the environment and bring them back to a fixed ``home'' location.
Flocking, on the other hand, is a process where robots that initially
move in different directions eventually agree to head in the same way.
Both case studies are commonly observed in biological systems and cover most of
the sources of complexity in MRSs.

This paper is structured as follows.
In Section~\ref{sec:features} we list a set of features commonly found in MRSs,
describing how they can represent a source of complexity during the
specification or analysis of these systems.
Section~\ref{sec:languages} introduces the considered languages and provides an
overview of their main features.
We then compare them on their ability to model the aforementioned features.
In Section~\ref{sec:casestudies} we describe our case studies and, if possible, 
we provide a basic implementation in each language, along with observations on
their respective advantages and limitations.
Section~\ref{sec:conclusions} contains our conclusions, as well as related and
future work.


\section{Common Features of Multi-Robot systems}\label{sec:features}
In this section we outline and categorize some common features of MRSs.
First of all, these systems are typically \emph{decentralized}, as they lack a
central unit of control: therefore there is no reliable way for a component to
obtain correct  information about the full state of the MRS.
Robots might also be free to join or leave the system at any time,
a feature known as \emph{open-endedness}. This may happen
deliberately (e.g. robots returning to a home location to charge their
batteries) or due to unexpected events, such as hardware failures.
Open-endedness complicates reasoning: for instance, robots leaving the system
create issues similar to those raised by failed processes in distributed
computing~\citep{Lamport1978a}.
Meanwhile, robots that join a MRS often need to gather information from
other components before they are able to cooperate; as another example, a 
robot might have to find alternative solutions when a collaborator leaves the
system. This calls for components with \emph{self-managing} capabilities, such
as self-configuration~\citep{Kephart2003}.
Furthermore, computational processes are \emph{distributed}, both physically and
logically, across the whole system.
Physical distribution, among other consequences, means that inter-process
communication may incur significant delays and possible failures.
On the other hand, logical displacement requires additional care to 
avoid well-known risks of concurrent programming, such as deadlocks and process
starvation.
As a further source of complication, often there are no temporal constraints to computation and interaction, as these systems may be partially or fully 
\emph{asynchronous}.
For instance, robots may take an arbitrary long time to send or read a message, and it is known that this can be the source of some fundamental problems in distributed computing, such as the distributed consensus~\citep{Fischer1985}.
Interaction in these systems is also characterized by
\emph{anonymity}, as it does not typically rely on identity.
Moreover, in a decentralized or open-ended system the whole concept of identity
is not easy to establish, and may even be irrelevant.
The ability to select partners according to their current task, or their
capabilities, can be more useful.
For instance, robots in a foraging swarm can perform \emph{recruiting} by
communicating the position of food to idle neighbours~\citep{Pitonakova2017}.
Such action only relies on the observed state of neighbours, and thus can
be performed also when agents are completely anonymous.

Due to the above mentioned features, in a MSR, the common interaction patterns of concurrent and distributed systems, such as point-to-point communication, shared memory, or synchronization, turn out to be inadequate.
Therefore, different solutions have been proposed which better fit to large, 
open-ended systems that do not rely on the concept of identity.
These include many-to-many communication (e.g. multicast or broadcast), or
even group-oriented interaction. A group-oriented network is composed of
groups of collaborating processes, and the message passing primitives do not target
individual processes, but whole groups~\citep{Birmanprocessgroupapproach1993}.
Indeed, robots that are popular in the MRS literature, like the
Kilobot~\citep{Rubenstein2012}, even lack hardware tools for
unicast or synchronous communication, making the aforementioned approaches the
only viable ones.

As the complexity of MRSs increases, there is a growing need for them to
react to new environmental conditions without human support.
This feature, known as \emph{adaptiveness}, is considered a necessity for
future computing systems as a whole~\citep{Kephart2003}, but is especially
attractive in the case of MRSs, since they are situated in a physical world
where a large number of unexpected situations may arise.
Adaptive behaviour can be found in various biological systems such as ant
colonies: when a source of food is found, ants collectively find an optimal path
from the nest.
When said path is disrupted, the colony is able to found a new, optimal one by
relying on a set of elementary actions performed by individual
ants~\citep{DorigoAntcolonyoptimization2006}.
This example also shows that system-wide adaptiveness can
be obtained even from simple actions by individual components.

Moreover, robots can be different from one another: their behavior, equipment
and capabilities may be \emph{heterogeneous}.
In principle, it is always possible to describe a system with differentiated behaviour as a homogeneous one, if the chosen formalism provides adequate
control-flow statements.
This approach can be acceptable for modelling mostly homogeneous swarms, but it is
insufficient when different groups of specialized robots are considered: in
fact, it introduces a significant overhead at various levels.
First of all, it greatly increases the complexity of the resulting
specification.
Because of that, very heterogeneous systems could become hard to understand and
maintain.
For instance, errors in the control flow may produce unwanted behaviour.
The need to differentiate behaviour at runtime may also negatively affect 
the performance of simulations and real-world implementations.
Finally, informal reasoning on system with a complicated control flow is difficult, as well as
verification through automated tools.

The \emph{large size} of the system may hinder the feasibility of
practical implementations, as it puts a high stress on the underlying runtime
environment and data structures.
Often, an individual behaviour that is acceptable when the number of robots is
small becomes unworkable as the population grows.
For instance, each robot can obtain a quite accurate view of a small system by
just exchanging messages with all the others.
But, due to the limited computational and networking capabilities of robots, this is usually not possible in the case of large MRSs.
A large size is detrimental to verification, and even simulation may become harder.
These effects are further complicated by \emph{non-determinism} and \emph{non-linearity}.
When the system is non-deterministic, multiple transitions are executable from a given
state, complicating the analysis.
Non-linearity, on the other hand, means that a local change may trigger a 
disproportionate, potentially system-wide effect.
As a consequence, simulation is not only more demanding, but also less
significant, as it might not spot critical, yet subtle cases where the system
fails.
The size of a system also plays a role in its classification.
For instance, nearly-homogeneous MRSs with a very large number of components 
typically fall into the \emph{swarm robotics} category.
A quantitative definition classifies a system of size $N$ as a swarm when
$10^2 < N <\!\!< 10^{23}$, with the rationale that ``Avogadro-large'' systems
are better treated with statistical
tools~\citep{BeniSwarmIntelligenceSwarm2005,Hamann2018}.
MRSs represent a more generic classification as they can be smaller in size but
more heterogeneous.

The presence of an \emph{environment} plays a critical role in MRSs, too. 
In fact, many applications of MRSs involve \emph{sensing} and \emph{actuation},
i.e. the gathering of data from, or manipulation of, the external environment.
It is often unfeasible to model these actions with the same tools used to
describe agents, since additional guarantees of synchrony, atomicity and
consistency must be provided which do not hold in agent-to-agent communication.
A similar observation has also been made in the more general context of
multi-agent
systems~\citep{DBLP:journals/fuin/WeynsH04,WeynsEnvironmentsmultiagentsystems2006}.

The manipulation of the environment can also work as a medium of indirect
interaction between robots.
This mechanism, known as \emph{stigmergy}, is often found in biological
systems~\citep{Grasse1959,Theraulaz1999}
and has some benefits over direct message passing.
For instance, it is inherently anonymous, as each agent simply react to changes
in the environment without knowing who caused them.
It is also considered a highly scalable solution~\citep{Heylighen2016}.
While inaccuracies in sensing and actuation can lead to lossy information
transfer, these advantages make stigmergic interaction attractive and widely
studied~\citep{Arkin1992,Werfel2005,Pitonakova2013}.



Additional pecularities of MRSs are strictly related to the knowledge of robots.
Each component  has only \emph{partial awareness} of the current state system it is operating in and possibly even of its own state.
For instance, robots can typically know the position of their neighbours, but not the one of robots that are farther away.
In open-ended systems, they might even not know the size of the system itself.
Even when robots are able to obtain information about the system, said
knowledge might be partial or become outdated by the time it is accessed.
This raises the problem of how to adequately represent knowledge and its
propagation among components, which is an important element to accomplish
complex coordination tasks~\citep{Pitonakova2017}.
As any kind of shared memory is unacceptable in large and distributed systems,
resorting to distributed data structures may be the only feasible approach.
However, the design of these structures must deal with the problems arising from
the extremely dynamic nature of MRSs, which can lead to integrity and
consistency problems.



\section{Languages}\label{sec:languages}


In this section we present a set of languages suitable for the specification
and analysis of MRSs, namely Buzz~\citep{Pinciroli2016}, ISPL~\citep{Lomuscio2017} and SCEL~\citep{DeNicola2014}.
Our selection is driven by their different nature, which makes them almost
orthogonal with respect to each other.
This allows us to better outline their respective strengths and drawbacks.
Buzz is oriented to real-world applications and provides a quite mature runtime
environment, including a reference virtual machine and a simulation platform
with an integrated physical engine.
Its similarities with popular general-purpose languages, such as Python,
JavaScript and Lua, can also be considered an advantage.
ISPL and its surrounding framework are specifically designed to support epistemic
logics, which enables reasoning on the knowledge of individuals or groups of
agents. Appropriate primitives are provided to model the interaction among agents 
and of agents with the environment.
Finally, SCEL is a process calculus that offers the possibility of naturally 
guaranteeing features such as anonymity and open-endedness.
This is made possible by its inherently group-oriented interaction primitives,
which rely on dynamic ensembles formed by taking into account the exposed features of components.
Thanks to its parametric semantics, the language can also be adapted to manage different  knowledge models.

\subsection{Buzz}
Buzz~\citep{Pinciroli2016} is a language for heterogeneous robot swarms.
It is designed as a core language that provides a few communication and
coordination primitives, and can be extended to suit the needs of the user.
For instance, it supports asynchronous communication with neighbours: each
component maintains a list of neighbours and can broadcast a key-value pair or
listen (in a non-blocking fashion) for a given key.
Swarms, i.e. dynamic ensembles of robots, are also a first-class abstraction in
Buzz.
Robots can join or leave an ensemble at runtime, and swarms can execute
arbitrary functions.
For instance, one could design a system where an ensemble periodically
broadcasts sensor data, while a second swarm receives these messages and uses
them to take decisions.

A distinctive feature of the language is the concept of virtual
stigmergies~\citep{Pinciroli2016a}.
Stigmergies represent a first-class abstraction of a shared knowledge base.
They are distributed key-value stores, replicated on all robots, where entries 
propagate or get overwritten based on their attached timestamps.
For instance, when two agents try to bind the same stigmergy key to different
values, there will be an initial phase where both entries will spread across the
system.
However, if the swarm is connected (i.e. each robot has at least one neighbour),
at some point the entry with the lower timestamp will stop propagating.
The newer entry, on the other hand, will continue spreading, eventually
overwriting the other one on the local copy of every component.
To avoid inconsistencies without resorting to a global clock, the mechanism
relies on Lamport timestamps~\citep{Lamport1978}.
This mechanism gives some guarantees over the eventual consistency of all
local copies of the virtual stigmergies. 

Buzz is very marginally concerned with embodiment, i.e. the fact that robots
are distinct entities situated in, and able to interact with, the physical
world~\citep{Brooks1991}.
Indeed, any kind of sensing and actuation, including the robot's own movements,
must be modelled by extending the language with appropriate functions.
This philosophy reduces the complexity of the language, but also leaves the
developer the responsibility of defining the semantics of the extensions.

The current implementation of Buzz requires all robots to have a unique
identifier, which is attached to all messages.
For instance, the tie-breaking protocol for virtual stigmergies reduces to a
comparison between IDs.
Moreover, all robots periodically broadcast their ID so that they
can use these messages to keep their neighbours list up-to-date.
These aspects might raise scalability issues, especially in dense swarms where
each robot could have tens of neighbours.
Buzz also assumes that a robot, upon receiving a message, can automatically
detect the position of the sender thanks to
\emph{situated communication} equipment~\citep{DBLP:conf/scai/Stoy01}.
As a consequence, there is no need for robots to declare their own position
in the message payload, making the maintenance of the neighbour list less
complex.
At the same time, situated communication devices currently face other
limitations: for instance, only robots that are in direct line-of-sight with
each other can exchange messages.
This might be a limiting factor in very cluttered environments.

MRSs defined through Buzz can be simulated on the ARGoS
platform~\citep{PinciroliARGoSmodularparallel2012}.
The user configures the number of robots and the Buzz script they execute by
editing an XML file. This file also describes the arena and its obstacles, the
spatial distribution of robots, and their equipment.
Even though the authors stress that all robots must execute the same Buzz
script, we were actually able to simulate a system where two groups of 
MarXbots~\citep{Bonani2010} execute slightly different scripts.
We observed that robots can communicate with each other through a virtual
stigmergy, even if they belong to different groups.
However, we also noticed that different agents may have the same identifier.
Even though our simulation behaved as expected, this violates the aforementioned
assumption on the uniqueness of IDs and therefore can lead to undesired
consequences.

\subsection{ISPL (Interpreted Systems Programming Language)}

The ISPL language~\citep{Lomuscio2017} is based on
\emph{interpreted systems}, a generalization of labeled transition systems where
multiple LTSs may synchronize on specific
actions~\citep{FaginReasoningknowledge1995}.
Each agent has a state (a set of user-defined variables) and a \emph{protocol}
that defines the actions it can perform given the current state.
Changes to the local state are encoded in a \emph{local evolution} function,
which takes into account both the current local state and the actions performed
by other agents.
As a consequence there are no explicit primitives for communication, the latter 
is encoded in the evolution of agents as the result of the synchronization on specific actions.
The state of agents is encapsulated: agents can only observe other agents'
actions, and eventually synchronize with them.
The environment is an exception, as it is a distinct agent whose variables may be fully
or partially observed by the other agents.
The synchronization mechanisms over actions can be used to model different
communication schemes involving an arbitrary number of agents.
Furthermore, specific actions can be defined so that they require the simultaneous
interaction of multiple agents and of the environment.

This way of modelling interaction, while flexible, also makes asynchronous 
interaction difficult to describe.
For instance, representing the delayed reception of a message requires 
the declaration of appropriate variables and evolutions within each agent.
Value-passing is difficult to describe as well.
In principle, agents should encode each possible value in a different action.
This approach may cause an increase in the complexity of agents, and does not
account for values over infinite domains.
Moreover, anonymity and open-endedness are not considered, as the system size
is fixed and transitions explicitly take into account other agents' actions.

These concerns have been partially addressed in the MCMAS-P
framework~\citep{KouvarosParameterisedverificationmultiagent2016},
where the size of system is parametrized.
The user only specifies the behaviour of each kind of agent in the system; in
the verification phase, concrete systems are instantiated by creating a fixed
number of agents for each role.
Thanks to cutoff techniques, verification of a property against a limited number
of concrete systems can be sufficient to prove that all concrete systems derived
from the same set of roles do satisfy the property.
However, finding a cutoff for an arbitrary property is, in general, an undecidable
problem.
Hence, the cutoff search algorithm of MCMAS-P is sound but incomplete.
Moreover, open-ended systems are still out of reach, as the size of each
concrete instantiation is fixed.


Verification is possible through model-checking of \emph{epistemic properties}
in the MCMAS framework.
An epistemic formalism is typically derived from an existing temporal logics by
adding modalities to ``reason about the knowledge of the agents in the
system''~\citep{Lomuscio2017}.
For instance, epistemic logics naturally allows to express properties such as
``All agents eventually know $\varphi$'' or ``Agent $i$ always knows
$\varphi$'', where $\varphi$ is another temporal or epistemic property.
MCMAS originally supported the ATLK language, an extension of Alternating
Temporal Logics (ATL)~\citep{Alur2002}.
A more recent implementation supports a significantly more expressive language,
LDLK~\citep{Kong2017}.
The same framework also provides support for interactive simulation, where the
user can choose an initial state of the system and manually select a sequence
of transitions to better understand the behaviour of agents.

\subsection{SCEL (Software Component Ensemble Language)}
SCEL~\citep{DeNicola2014,DeNicola2015} is a formal language for the description
and verification of collective adaptive
systems~\citep{Hillston2014}.
In order to capture the highly dynamic nature of this class of systems, it
naturally supports concepts such as open-endedness and anonymity.


Communication in SCEL is deeply related to the concept of knowledge
repositories.
A knowledge repository is a container of knowledge items.
The nature of repositories and items is not specified, and the SCEL semantics
is parametric with respect to their semantics.
While most works use tuple spaces~\citep{Gelernter1985}, other kinds of
repositories have been proposed.
Soft constraint programming~\citep{Schiex:1995:VCS:1625855.1625938},
for instance, can be integrated into SCEL by defining repositories as constraint
stores~\citep{Montanari2015}.
Each component is equipped with a set of \emph{attributes}, which are named
values exposed to the whole system.
Attributes and their values are stored in the knowledge repository of the
component, thus they can change during the evolution of the system.
The set of exposed attributes, known as \emph{interface}, can be dynamic as
well.


Communication is achieved through the manipulation of said repositories.
In addition to inserting items (via the \textbf{put} action), a component can
read or withdraw an item that matches a specified \emph{pattern}, or template
(via the \textbf{qry} and \textbf{get} actions, respectively).
All operations are either point-to-point or \emph{attribute-based}, i.e
involving only those components that satisfy a given predicate over their
exposed attributes.
This leads to a high degree of anonymity, as components do not need to know the 
identity of interaction partners, nor to expose their own.
Components are able to manipulate their own repository by using the
special $\Self$ identifier.
The knowledge repository is also an abstraction layer over sensing and
actuation.
For instance, the intention of an agent to move towards a destination $d$ could
be represented by putting a $(\Quote{moveTo}, d)$ item into its own repository.
It is assumed that another process will retrieve this tuple, drive the agent's
motors accordingly, and potentially announce the result of the operation by
inserting another tuple in the repository.
Notice that the \textbf{put} action is the only non-blocking one.
The blocking nature of \textbf{qry} and \textbf{get} is useful to implement
various reactive patterns and to guarantee processes synchronization.
Guards, for instance, are naturally implemented by waiting until a specific item
can be withdrawn from the component's own repository.
When the semantic of repositories is similar to that of tuple spaces,
\emph{generative communication} patterns can be easily
applied~\citep{Carriero1989,Carriero1994}.

Implementing a runtime environment that respects the SCEL operational semantics
is not trivial.
The jResp implementation provides three options.
The first one relies on a centralized message broker, which may be unacceptable
in  scenarios where full decentralization is needed.
Alternatively, messages and predicates could be broadcasted in a bus-like
topology.
In this case, receivers accept or reject a given message after evaluating the
associated predicate over their own interface.
This solution is completely decentralized, but evidently does not scale well
with the size of the system.
The third option is a peer-to-peer topology based on the Scribe
protocol~\citep{CastroScribelargescaledecentralized2002}.

With jResp it is also possible to simulate the computational aspects of a SCEL
system.
However, physical simulation (such as the one offered by Buzz through ARGoS) is 
currently unavailable.
Multiple verification approaches exist for systems specified in SCEL and
its derivatives.
The subset of SCEL without policies, known as SCELight, can be directly
translated to Promela, thus allowing for model-checking through the SPIN
tool~\citep{DeNicola2014a}.
The MISSCEL implementation~\citep{Belzner2014}
enables simulation and logical model-checking within
the \textsc{Maude} framework.
Furthermore, \textsc{MultiVeStA} can be used to verify systems through 
statistical model-checking, a technique based on checking  formulae satisfaction against a finite number of executions.
As a consequence, it can only provide a statistical evidence that the required
property is satisfied. On the other hand, it is highly parallelizable and can
provide insight on systems that are too large to be formally
verified~\citep{Legay2010}.

\subsection{Comparison}
We summarize  in Table~\ref{tab:comparison} our findings about the three formalisms.

\begin{table}[h!]
    \caption{Comparison of MRS support of the languages we considered.}
    \label{tab:comparison}
    \centering

    \begin{tabular}{p{2.8cm}p{4.3cm}p{4.45cm}p{4.45cm}}
    \hline

    \hline
     & \textbf{Buzz} & \textbf{ISPL} & \textbf{SCEL} \\
    \hline
        \raggedleft\emph{Open-endedness}
            & No
            & No
            & Yes\\
        \raggedleft\emph{Asynchrony}
            & Yes
            & No
            & Yes\\
        \raggedleft\emph{Anonymity}
            & Medium (\texttt{neighbors})
            & Low
            & High\\
        \raggedleft\emph{Heterogeneity}
            & Low
            & High
            & High\\
        \raggedleft\emph{Communication}
            & Ranged broadcast
            & Multicast
            & Multicast\\
        \raggedleft\emph{Knowledge representation}
            & \raggedright Local variables, virtual stigmergies
            & \raggedright Local and environmental variables
            & \raggedright\arraybackslash Parametric (e.g. tuple spaces) \\
        \raggedleft\emph{Environment}
            & No
            &  Yes
            & Yes (as an additional component)\\
        \raggedleft\emph{Semantics}
            & Reference implementation
            & \raggedright Formal (Kripke structures)
            & Formal (SOS)\\
        \raggedleft\emph{Analysis}
            & Physics-based simulation (ARGoS)
            & \raggedright Simulation, Model-checking (\textsc{MCMAS})
            & \raggedright\arraybackslash Simulation (jResp); 
            Model-checking (SPIN, \textsc{Maude}); statistical model checking (\textsc{Vesta})\\
    \hline

    \hline
    \end{tabular}
\end{table}

We found SCEL to be the only language with the capability to represent systems
of dynamic size.
Its syntax contains a \texttt{new} keyword that allows components to
``spawn'' additional agents.
This can be crucial to naturally specify fully open-ended MRSs.

ISPL is different from both Buzz and SCEL in that it enables explicit
specification of the environment.
It is not clear whether implementing an environment on top of Buzz would be
possible, as its primitives are fully asynchronous and oriented to concrete
robots.
Extending the language seems the most appropriate approach.
Meanwhile, using one or more SCEL components to model an environment could be
viable, since the language provides both asynchronous and synchronous
mechanisms,
and a more flexible representation of knowledge.
ISPL has no primitives for asynchronous interaction, nor value-passing.
This means that replicating asynchronous features, while possible in principle,
would be quite complex and might negatively affect verification times.
The different approaches to communication are also reflected in the support for
anonymity. In ISPL, anonymity is low as interaction happens through
synchronization with specific agents: the mediation of the environment can
represent a solution, like in our flocking example below, but adds complexity to the 
specification.
Buzz offers the possibility of broadcasting or listening messages among neighbours,
but the dependence on unique IDs contrasts with full anonymity.
In SCEL, by contrast, attribute-based actions are inherently anonymous, as
sender and receivers can communicate without any specific information on each
other.

As regards the analysis of specified systems, to the best of our knowledge Buzz
is the only language equipped with a physics-based simulation environment.
This could be important to study the behaviour of agents under conditions that
they could face in the real world.
Meanwhile, Buzz offers little support for formal verification.
On the other hand, no language except SCEL provides documented support
for statistical model checking, which could be a useful tool for large MRSs.
However, given that MCMAS can already compute traces from a given ISPL system, 
extending ISPL to support this technique seems relatively straightforward.


\section{Case studies}\label{sec:casestudies}

\subsection{Foraging}

Foraging is considered a canonical case study in the literature related to MRS
and robotic swarms, since it can be used to model many kinds of scenarios,
such as ``waste retrieval'' and ``search and rescue''~\citep{Brambilla2013}.
Specifying a foraging MRS can show the capabilities of the chosen specification 
language with respect to different features of these systems, such as the
representation of the agents' knowledge and their interaction with the
environment through sensors and actuators.


\textbf{Buzz}. As Buzz lacks a notion of environment as well as
synchronous communication primitives, food items have to be implemented as
components, and specific protocols must be set up to limit inconsistencies.
The system consists of into two \emph{swarms}: food items and forager agents.
Foragers perform a random walk and repeatedly broadcast a pick-up request to all
neighbours.
Food items wait for pick-up requests and if the requesting forager is close enough, they
respond with its id to signal that they have been collected successfully.

\begin{lstlisting}[
    language={[5.0]Lua}, 
    showstringspaces=false,
    basicstyle=\tiny\ttfamily,
    commentstyle=\scriptsize\itshape,
    otherkeywords = {math, pose, stigmergy, var}]
function food_listen(vid, value, rid) {
  # Check if the robot is closer than 50 cm
  d = neighbors.get(rid).distance
  if (d < 50) {
    neighbors.broadcast("response", rid)
    # Stop responding to other foragers
    neighbors.ignore("pick_up")
  }
}

function robot_listen(vid, value, rid) {
  if (value == id) {
    log(id, ": picked up ", rid)
  }
}

function init() {
  # Robots with id = 0, 4, 8 etc. are food items
  foodSwarm = swarm.create(1)
  foodSwarm.select(id % 4 == 0)
  # All the others are foragers
  foragerSwarm = foodSwarm.others(2)
  if (foodSwarm.in()) {
      neighbors.listen("pick_up", food_listen)
  } else {
      neighbors.listen("response", robot_listen)
  }
}

function step() {
  if (foragerSwarm.in()) {
    # Only foragers execute this block
    # Random walk (sets linear and angular velocity)
    gotop(5, math.rng.uniform(-3.0, 3.0))
    neighbors.broadcast("pick_up", id)
  }
}
\end{lstlisting}

We encode the individual behaviour inside two standard Buzz functions.
The first one is \texttt{init()}, which will be executed only once, after the
agent has been created.
The function \texttt{step()}, instead, is repeatedly executed by each robot
until the experiment terminates.
One might think of stopping food items once they signal their availability to
one of the foragers, but it appears that the execution of function
\texttt{step()} cannot be blocked through any statement provided by the
language.

This solution has still some limitations. For instance, it is still possible for
two foragers to pick up the same food item. Indeed, if a food item receives two \texttt{pick-up} messages, 
it could perform the listener function \texttt{food\_listen} twice and send two different
\texttt{response} messages.



\textbf{ISPL}. 
Let us assume that the arena is a two-dimensional grid of size $10 \times 10$.
We use the Environment agent to keep track of the position of food items and to
recorder whether they have been collected or not.
This information needs to be stored in \emph{observable variables}, since
foraging robots need to access it.
We also define an internal variable with the number of found items, which will
be used for verification purposes.
Robots, on the other hand, are agents with a position.
We only give the specification of \texttt{Robot1}, as the only difference
between foragers is their identifier.

\begin{lstlisting}[
    language={[5.2]Lua}, 
    showstringspaces=false,
    basicstyle=\tiny\ttfamily,
    commentstyle=\scriptsize\itshape,
    linerange={1-13,26-33, 53-54},
    otherkeywords = {boolean, Obsvars, Vars, Agent, if}]
Agent Environment
  Obsvars:
    food1X : 1 .. 10;
    food1Y : 1 .. 10;
    food1 : boolean;
    
    food2X : 1 .. 10;
    food2Y : 1 .. 10;
    food2 : boolean;
  end Obsvars
  Vars:
    foundItems : 0 .. 2;
  end Vars
  -- ...
end Agent

Agent Robot1
  Vars:
    x : 1 .. 10;
    y : 1 .. 10;
  end Vars
  -- ...
end Agent
\end{lstlisting}

We define the protocol and evolution functions so that foragers perform a
random walk and can only pick up an item when they are in its same position, and
it has not been collected yet.
When an item is collected, the environment updates the corresponding boolean
variable and increments the internal counter.

\begin{lstlisting}[
    language={[5.2]Lua}, 
    showstringspaces=false,
    basicstyle=\tiny\ttfamily,
    commentstyle=\scriptsize\itshape,
    linerange={17-21,25-25,34-44,46-52},
    morekeywords={Action, Actions, Protocol, Evolution, if, end}
    ]
  -- Environment
  Evolution:
    food1 = false and foundItems = foundItems+1 if
      (Robot1.Action = PickFood1);
    -- Repeat for all robots and food items
  end Evolution
  -- Robots
  Actions = {Up, Down, Left, Right, PickFood1, PickFood2};
  Protocol:
    y < 10 : {Up};
    y > 1 : {Down};
    x < 10 : {Right};
    x > 1 : {Left};
    Environment.food1 = true and 
      x = Environment.food1X and 
      y = Environment.food1Y : {PickFood1};
    -- Repeat for all food items
  end Protocol
  Evolution:
    x = x+1 if (Action = Right);
    x = x-1 if (Action = Left);
    y = y+1 if (Action = Up);
    y = y-1 if (Action = Down);
  end Evolution
\end{lstlisting}

Finally, we specify some constraints on the initial state of the system: namely,
all items are available and the \texttt{foundItems} counter is set to 0.
As we specify no restriction on the positions of robots and items, their value
will be fully non-deterministic.

\begin{lstlisting}[
    language={[5.2]Lua}, 
    showstringspaces=false,
    basicstyle=\tiny\ttfamily,
    commentstyle=\scriptsize\itshape,
    linerange={83-87},
    morekeywords={InitStates}
    ]
InitStates
  Environment.food1 = true and
  Environment.food2 = true and
  Environment.foundItems = 0;
end InitStates
\end{lstlisting}

Alternatively, one could model food items as components.
Parametrised versions of this approach are available in the
literature~\citep{KouvarosParameterisedverificationmultiagent2016}, but to our
knowledge these models do not take into account the physical location of robots
and items.

\textbf{SCEL}.
Let us assume knowledge repositories to be tuple spaces.
The implementation we describe is based on more complex examples, related to
search and rescue operations, available in the existing literature on
SCEL~\citep{DeNicola2014,DeNicola2015}.

We can model both foraging robots and food items as SCEL components.
Each forager exposes its own position \emph{pos}, the task it is performing
(initially all robots are \emph{idle}), and the range of its sensor.
Each food item runs the same process $P_{\mathit{food}}$:

\begin{align*}
P_{\mathit{food}} \triangleq\; &
\Put{\Quote{food}, \This.pos}
{\left(\textit{task} = \Quote{idle}
\land \Vert pos - \This.pos \Vert \leq range \right)}
.P_{\mathit{food}}\\
& +\\
& \Qry{\Quote{found}}{\Self}.\Nil
\end{align*}

This means that the food item will repeatedly communicate its own position to
all idle foragers that are closer than the range of their own sensor.
This process will terminate when the item discovers a 
$(\textit{``found''})$ tuple into its own knowledge repository.
($\Nil$ denotes the inactive process).
Before describing the behaviour of foragers, let us assume that each food item
initially has a $(\textit{``lock''})$ tuple in its repository.
We will use it to ensure that only one forager is able to collect the item.

Foragers alternate between the idle and working states.
In the idle state, they just perform a random walk until they sense a food item.
In that case, they change their \emph{task} attribute accordingly and move
towards the food source.

\begin{align*}
P_{\mathit{idle}} \triangleq\; &
    \Get{\Quote{food}, \textit{?f}}{\Self}.
    P_{\mathit{work}}(f) \;+\; \Put{\Quote{randomWalk}}{\Self}
    .P_{\mathit{idle}}\\
P_{\mathit{work}} (\mathit{food}) \triangleq\; &
    \Put{\Quote{task}, \Quote{work}}{\Self}.\\
    & \Put{\Quote{moveTo}, \mathit{food}}{\Self}.
    \Qry{\Quote{reached}, \mathit{food}}{\Self}.(\\
    &
    \begin{array}{ll}
    & \Get{\Quote{lock}}{(pos = food)}.
    \Put{\Quote{found}}{(pos = food)}.\\
    & \Put{\Quote{task}, \Quote{idle}}{\Self}.P_{idle}\\
    & + \\
    & \Put{\Quote{task}, \Quote{idle}}{\Self}.P_{idle})
    \end{array}
\end{align*}

The syntax $\left(\Quote{food}, \textit{?f}\,\right)$ denotes a template.
In this case, the template is matched by all two-element tuples where the first
element is $\Quote{food}$:
when such a tuple is found, it is removed from the repository and its second
element is bound to the variable $f$.
Like in most programming languages, $P_{\mathit{work}}(f)$ denotes a parametric
invocation, where the actual parameter $f$ is bound to the formal parameter
$\mathit{food}$.
As stated in Section~\ref{sec:languages}, we use special tuples, such as
\emph{moveTo} and \emph{reached}, to represent the start and stop of a movement.
Notice that updating the value of an attribute, such as \emph{task}, needs no
additional primitives, as it just can be obtained by manipulating items in the
local repository.
As said above, to pick up an item, a robot must first withdraw its \emph{lock}
tuple.
If this is not possible, it means that another forager has already picked up
the item: hence the robot simply turns back to the idle state.

\subsection{Flocking}

Flocking is an example of emerging behaviour where agents starts from a state of incoherent motion, but eventually agree to move in the same direction.
This case study is a basic instance of a consensus problem, which is fundamental
for many cooperative tasks~\citep{Valentini2017}.

Except for cases where the language provides better suited primitives, we will
address this problem by specifying some variation on the \emph{voter model}.
In a voter model, agents are seen as nodes in a graph that initially have 
different \emph{opinion} about what to choose among a 
finite number of possibilities~\citep{Liggett2005}.
Furthermore, each agent can observe and copy the opinion of a random neighbour.
A voter model can either converge (i.e. all agents eventually adopt the same
opinion) or oscillate, based on a number of factors, such as the initial 
distribution of opinions, the topology of the graph, etc.

\textbf{Buzz}. We can easily adopt a virtual stigmergy to make the swarm agree
on a direction (Table~\ref{tab:vstig}).

\begin{table}[!h]
    \caption{Flocking in Buzz with virtual stigmergies.}
    \label{tab:vstig}
    \centering
\begin{lstlisting}[
    language={[5.0]Lua}, 
    showstringspaces=false,
    basicstyle=\tiny\ttfamily,
    commentstyle=\scriptsize\itshape,
    otherkeywords = {math, pose, stigmergy, var}]
function init() {
    math.rng.setseed(id*id)
    my_yaw = math.rng.uniform(0, 2.0 * math.pi)
    vstig = stigmergy.create(1)
    vstig.put("yaw", my_yaw)
}

function step() {
    var cur_yaw = pose.orientation.yaw % (2. * math.pi)
    var err = vstig.get("yaw") - cur_yaw
    control(err) 
}
\end{lstlisting}
\end{table}

In the \texttt{init()} function we use the squared ID of the agent as a seed for
the pseudo-random number generator (PRNG): this is needed to make each agent
behave differently.
We then generate a random value for the direction and put it into a virtual
stigmergy.

At each execution step, robots retrieve the direction from the stigmergy by
calling the \texttt{vstig.get()} function, and compute the current error (i.e
the difference between the desired yaw angle and the current one).
The \texttt{control()} function then rotates the robot if the error is too big,
and makes it move forward otherwise.
An elementary implementation of such a function is the following:

\begin{lstlisting}[
    language={[5.0]Lua},
    alsolanguage={Python},
    showstringspaces=false,
    basicstyle=\tiny\ttfamily,
    commentstyle=\scriptsize\itshape,
    otherkeywords = {math, pose, stigmergy, var}]
function control(err) {
    if (math.abs(err) > 0.5) {
        sign = err / math.abs(err)
        gotop(0, sign * 2) # Rotate (+/-  2 rad/s)
    }
    else {
        gotop(3, 0) # Move forward (3 cm/s)
    }
}
\end{lstlisting}


Each time a robot reads a value from the stigmergy, the Buzz virtual machine
automatically asks its neighbours to confirm whether its local value is
up-to-date.
Neighbours either use this information to update their own local copies or to
reply with a more recent value.
This mechanism makes it easier for all robots to converge to a common value,
even when parts of the swarm become temporarily disconnected from the rest.

A basic voter model can also be described in Buzz.
We can use the neighbour communication primitives to enable each robot
to broadcast its chosen direction among its neighbours.
Periodically, robots listen to the direction of a neighbour and change their
own accordingly (Table~\ref{tab:voter}).

\begin{table}[!h]
    \caption{Flocking in Buzz: a voter-model approach.}
    \label{tab:voter}
    \centering
\begin{lstlisting}[
    language={[5.0]Lua},
    showstringspaces=false,
    basicstyle=\tiny\ttfamily,
    commentstyle=\itshape,
    otherkeywords = {math, neighbors, pose, stigmergy, var}]
t = 0

function init() {
    math.rng.setseed(id*id)
    my_yaw = math.rng.uniform(0,2) * math.pi
    neighbors.listen("yaw", function(vid, value, rid) {
        if(value != my_yaw and (t % 20) == 0) {
            my_yaw = value
            log(id, " change to ", my_yaw)
        }
    })
}

function step() {
    t = t + 1
    var cur_yaw = pose.orientation.yaw % (2. * math.pi)
    var err = my_yaw - cur_yaw
    control(err)
    neighbors.broadcast("yaw", my_yaw)
}
\end{lstlisting}
\end{table}

Here the condition \texttt{(t \% 20) == 0} means that robots can only
attempt to change their opinion once every 20 time steps.
By altering the condition it is possible to simulate other kinds of models.
For instance, the waiting time of each robot could follow an
exponential~\citep{Cox1989} or power-law~\citep{Takaguchi2011} distribution.
The effects of \emph{zealots}, i.e. agents that never change their
opinion~\citep{Mobilia2007}, could also be studied.

\textbf{ISPL}.
We decided not to rely on explicit communication between robots.
To do so, we should create appropriate protocol and evolution rules for each
pair of agents: therefore the size of the specification would grow quadratically
with the number of robots.
Our implementation, again, takes advantage of the \texttt{Environment} agent.
Each robot starts with a random direction stored in its state.
At any moment, a robot can move in its stored direction, or it can watch and
imitate the direction of the last robot that moved.

\begin{lstlisting}[
    language={[5.2]Lua}, 
    showstringspaces=false,
    basicstyle=\tiny\ttfamily,
    commentstyle=\scriptsize\itshape,
    linerange={1-5,11-13,25-2, 25-35, 40-41, 45-48, 57-58, 62-63},
    morekeywords = {boolean, Actions, Protocol, Obsvars, Vars, Agent, if, Evolution}]
Agent Environment
  Obsvars:
    dir : {Up, Down, Left, Right}; 
  end Obsvars
  -- ...
  Evolution:
    dir = Up if (Robot1.Action = Up);
    -- Repeat for all robots and directions

  end Evolution
end Agent

Agent Robot1
  Vars:
    x : 1 .. 10;
    y : 1 .. 10;
    dir : {Up, Down, Left, Right};
  end Vars
  Actions = {Up, Down, Left, Right, Watch};
  Protocol:
    dir = Up and y < 10: {Up, Watch};
    -- Similarly for other directions
    Other: {Watch};
  end Protocol
  Evolution:
    dir = Environment.dir if (Action = Watch);
    y = y+1 if (Action = Up);
    -- Similarly for other directions
  end Evolution
end Agent
\end{lstlisting}

We define the protocol so that robots cannot move in their chosen
direction if they are are on the edges of the arena.
For instance, an agent at position $(3,10)$ cannot go in the \texttt{Up}
direction. In these cases, robots can only perform the \texttt{Watch} action.
We encode this with the special condition \texttt{Other}, which holds in all
the states that do not match the other protocol rules.
In this case we were able to verify that two robots will always agree to move in
the same direction, by checking a property of the form
$\textbf{AF }\mathit{consensus}$ (for all possible executions, eventually the
\emph{consensus} proposition will hold).
However, when there are three or more robots, the MCMAS model checker is able to
find cyclic traces where consensus is never achieved.

We can slightly alter the description of robots, so that they can move across an
edge and get to the opposite side of the grid.
For instance, a robot at $(1,4)$ will be able to move left and reach $(10, 4)$.
In other words, we consider a toroidal, rather than square, arena.
With these changes we are unable to prove global properties about consensus,
even for the two-robot case.

\begin{lstlisting}[
    language={[5.2]Lua}, 
    showstringspaces=false,
    basicstyle=\tiny\ttfamily,
    commentstyle=\scriptsize\itshape,
    morekeywords = {boolean, Actions, Protocol, Obsvars, Vars, Agent, if, Evolution}]
Protocol:
  dir = Up : {Up, Watch};
  -- ...
end Protocol
Evolution:
  y = 1 if (Action = Up and y = 10);
  y = y+1 if (Action = Up and y < 10);
  -- ...
end Evolution
\end{lstlisting}

While ISPL cannot express voter models in a natural way, an ad-hoc version of
the language (ISPL-OFP) has been implemented to model and verify an array of
opinion formation protocols~\citep{Kouvaros:2016:FVO:2936924.2937099}.

\textbf{SCEL}.
Modelling a basic voter model in SCEL is simple.
Each component exposes its current position and direction, and uses the
\textbf{qry} action to copy the direction of a neighbour.
Due to the semantics of the action, a neighbour will be selected
nondeterministically among the components that satisfy an attribute-based
predicate.
In this example the predicate relies on an additional attribute, storing the
communication range of the component, and only targets neighbours exposing
a different direction.

\begin{align*}
P \triangleq\; &
    \Qry{\Quote{direction}, \textit{?d}}
    {\left(\Vert \mathit{pos} - \Self.\mathit{pos} \Vert \leq \Self.\mathit{range}
    \land \mathit{direction} \neq \Self.\mathit{direction}\right)}.\\
    & \Put{\Quote{direction},d}{\Self}.P
\end{align*}

We can further refine this behaviour by introducing logical
clocks~\citep{Lamport1978}, obtaining a protocol similar to the one of Buzz
based on virtual stigmergies.
Suppose that each component has a $(\Quote{time}, 0)$ tuple.
This value is also exposed as the \emph{time} attribute of the component.
We use this attribute to the \textbf{qry} predicate to ignore out-of-date
neighbours.
We also attach a timestamp to \emph{direction} tuples.
Whenever a component finds a neighbour with a timestamp $t$ higher than its own,
it sets its own clock to $t+1$.
A similar approach can be taken to divide the computation into ``rounds'' and
solve more complex problems, such as distributed graph coloring, see e.g.,~\citep{Alrahman2017}.

\begin{align*}
P \triangleq\; &
    \Qry{\Quote{direction}, \textit{?d}, \textit{?t}}
    {\left(\Vert \mathit{pos} - \Self.\mathit{pos} \Vert \leq \Self.\mathit{range}
    \land \mathit{time} \geq \Self.\mathit{time}\right)}.\\
    & \Put{\Quote{time}, t+1}{\Self}.\\
    & \Put{\Quote{direction}, d , t+1}{\Self}.P
\end{align*}


\section{Conclusions, related and future work}\label{sec:conclusions}

In this work, we have described a number of features typically found in
multi-robot systems.
These traits can make MRSs hard to design, implement and reason about.
We have presented a selection of languages, showing how specific linguistic
primitives can help in the specification of such systems.
We also compared the languages considering the tools that have been provided to support analysis 
of their systems.
To better understand the strengths and weaknesses of each language, we
considered two case studies that are popular in the MRS literature, and
provided basic implementation in the surveyed languages.
We used these implementations to show how specific abstractions provided by the
languages facilitates  modelling non-trivial behaviours.
For instance, the presence of synchronous operations is vital for scenarios
where the interaction with the environment is predominant.
At the same time, we have seen that group-oriented forms of interaction,
either among neighbours or based on the more general framework of
attribute-based communication, allow for a more natural specification of
non-trivial cooperation between agents.
Our work is by no means an exhaustive review of the state of the art.
We focused on choosing a meaningful set of features to allow an effective
qualitative comparison of existing languages. Additional languages and
frameworks could be the subject of further investigation.
Describing and clarifying the factors that make MRSs distinctively 
challenging is also an important first step towards the specification of new
MRS-oriented languages, which is another possible direction of research.

\textbf{Related work.} A number of language surveys can be found in the field of Multi-Agent Systems~\citep{DBLP:conf/woa/MascardiDA05,Bordini2006,Feraud2017}.
As MASs are a superset of multi-robot systems, said surveys may lack
detail and miss specific features highlighted in our work.
Surveys of robotics languages and platforms are also
available~\citep{Kramer2007,Nordmann2014}, but to the best of our knowledge they
do not address the sources of complexity presented in this overview.
Peculiar traits of complex systems in general, such as emergence, are also the
subject of a substantial amount of
research~\citep{Heylighen1989,Barabasi1999,Odell2002}.
The literature also provides taxonomies for specific aspects found in MRSs, such
as coordination~\citep{Yan2013} and task allocation~\citep{Gerkey2004}.

\textbf{Future work.} This work is for us instrumental to design a new language for multi-agent systems by building on the lesson learned from the three languages we have surveyed. The language we are aiming at should make it possible an intuitive design of local specifications and automated analysis of global properties and emerging behaviours. We will aim at a language that combines stigmergic interaction of Buzz with the attribute-based communication of SCEL,  whose  agents will interact by manipulating and asynchronously propagating their limited share of knowledge. The language will be equipped with a formal semantics to enable automatic verification of logical properties by building on  tools and methods developed for ISPL.

\bibliographystyle{frontiersinSCNS_ENG_HUMS} 
\bibliography{biblio}

\end{document}